\documentclass[a4paper,12pt]{article}
\usepackage[cp1251]{inputenc}
\usepackage{epsfig}
\usepackage[english]{babel}
\usepackage{amssymb}

\begin{document}
\title{Period tripling accumulation point for complexified H\'{e}non map}
\author{O.B. Isaeva, S.P. Kuznetsov}
\date{}
\maketitle\begin{center} \emph{Institute of Radio-Engineering and
Electronics of
RAS, Saratov Branch, \\ Zelenaya 38, Saratov, 410019, Russia \\
E-mail: IsaevaOB@info.sgu.ru}\end{center}

\begin{abstract}
Accumulation point of period-tripling bifurcations for
complexified H\'{e}non map is found. Universal scaling properties
of parameter space and Fourier spectrum intrinsic to this critical
point is demonstrated.
\end{abstract}

It is known that complexification of real 1D logistic map
\begin{equation}\label{log}
z_{n+1}=\lambda-z_{n}^{2},
\end{equation}
where $\lambda,z\in\mathbb{C}$ leads to the origination of the
Mandelbrot set at the complex parameter $\lambda$
plane~\cite{Peitgen} and a number of other accompanying phenomena.
So, except for the transition to chaos through cascade of
period-doubling bifurcations, intrinsic to the real logistic
map, other bifurcation sequences are possible. For example, the
universal properties of transition to chaos through
period-tripling cascade is studied in paper of
Golberg-Sinai-Khanin~\cite{Golberg}. Corresponded critical point
(GSK point) is situated at
\begin{equation}\label{lam}
\lambda_{c}=0.0236411685377+0.7836606508052 i
\end{equation}
and characterized by following critical indexes~\cite{Golberg},
namely, by critical multiplier $\mu_c$, scale factor $\alpha$ and
parameter scaling constant $\delta$ (see~Fig.1(a)):
\begin{equation}\label{mu}
\mu_c=-0.47653179-1.05480867 i,
\end{equation}
\begin{equation}\label{al}
\alpha=-2.0969+2.3583 i,
\end{equation}
\begin{equation}\label{del}
\delta=4.6002-8.9812 i.
\end{equation}

Opportunity  for realization of the phenomena, characteristic for
the dynamics of complex maps (Mandelbrot set etc.) at the physical
systems seems to be interesting problem~\cite{Beck}. In the
context of this problem the following question is important: Does
phenomena of dynamics of the 1D complex maps like classic
Mandelbrot map~(\ref{log}) survive for the two-dimensional maps.
For example, from the point of view of possible physical
applications, more realistic model rather than logistic map, is
the H\'{e}non map
\begin{equation}\label{hen}
z_{n+1}=f(z_{n},w_{n})=\lambda-z_{n}^{2}-d\cdot w_{n},\qquad
w_{n+1}=g(z_{n},w_{n})=z_{n}.
\end{equation}
In the real variable case the system~(\ref{hen}) is 2D invertible
map and, hence, can be realized as Poincare cross-section of flow
system with three dimensional phase space -- minimal dimension,
providing opportunity of nontrivial dynamics and chaos. H\'{e}non
map is suitable for modelling of the chaotic dynamics of the
generator with non-inertial nonlinearity, dissipative oscillator
and rotator with periodic impulse driving force
etc.~\cite{Kuznetsov}. Moreover, H\'{e}non map expresses the
principal properties of large class of differential systems.

Let us complexify the map~(\ref{hen}) in a such way that
$z,w,\lambda\in\mathbb{C}$, $d\in\mathbb{R}$. According to the
work~\cite{Isaeva,Isaeva1}, complexified H\'{e}non map can be
reduced to the two symmetrically coupled real H\'{e}non maps and
can be realised at the physical experiment~\cite{Isaeva2}.

Let us remark that with $|d|<1$ the H\'{e}non map is dissipative
system, with $|d| \rightarrow 1$ -- it is area-preserving map, and
with $d \rightarrow 0$ it corresponds to complex 1D quadratic
map~(\ref{log}), which describes the universal scenario of
transition to chaos through period-tripling bifurcations. In
present work we aim to be convinced of existence of GSK critical
point for the H\'{e}non map with $1>|d|\neq0$, to find this point
and to demonstrate intrinsic to it scaling characteristics.

Let us shortly explain essence of the numerical procedure for the
calculation of the GSK point. From the work~\cite{Kuznetsov1}, the
simple and effective method of calculation of the critical point
of transition to chaos though period-doubling bifurcations for 1D
maps and also for the more general systems is known. We have
generalized this method to the case of GSK critical point and have
applied it to the 2D complex H\'{e}non map. The method is based on
the fact, that at the GSK point the infinite number of the
unstable cycles with tripling periods must exist, and multipliers
of these cycles must tend to the universal constant $\mu_c$.
Therefore, for the finding of the point $\lambda_c^{d \neq 0}$ it
is enough to calculate  with fixed parameter $d$ the point at the
$\lambda$ plane, satisfying to these conditions. Therefor it is
necessary to solve the system of nonlinear complex equations:
\begin{equation}\label{eqh1}
f^{N}(z_N,w_N)=z_N,\quad g^{N}(z_N,w_N)=w_N,\quad \mu
(z_N,w_N)=\mu_c
\end{equation}
concerning the cycle elements $z_N$, $w_N$, where $N=3^k$
($k\rightarrow\infty$) and parameter $\lambda$. The system of
equations~(\ref{eqh1}) can be solved numerically by Newton
procedure. The multiplier $\mu(z_N,w_N)$ can be found as maximal
by absolute value eigenvalue of the Jacobi matrix ${\rm{\bf
J}}(z_N,w_N)$, where
\begin{equation}\label{jacobi1}
\mathbf{J}(z,w)={\left( {{\begin{array}{*{20}c}
 {{\frac{{\partial f^{N}(z,w)}}{{\partial z}}}} \hfill &
{{\frac{{\partial g^{N}(z,w)}}{{\partial z}}}} \hfill \\
 {{\frac{{\partial f^{N}(z,w)}}{{\partial w}}}} \hfill &
{{\frac{{\partial g^{N}(z,w)}}{{\partial w}}}} \hfill \\
\end{array}} } \right)}.
\end{equation}

Let us mark, that for the best convergence of a Newton method,
expediently to use inheriting starting conditions. At the each
consequent step of calculations, as a starting position of the
critical point the value of the parameter $\lambda$ obtained at
the previous step is used. As starting position of the elements of
cycles the values obtained according to the universal scaling
properties of the phase space at the critical GSK point are used.
First elements of different cycles selected properly should obey
the equations
\begin{equation}\label{elcyc}
z_N=z_c+C_1\alpha^{-N},\qquad w_N=w_c+C_2\alpha^{-N},
\end{equation}
where $z_c$ and $w_c$ -- are the constants, which correspond to
the scaling center at the phase space, and $C_1$, $C_2$ --are some
complex coefficients. In the case of the 1D complex map, i.e. with
$d=0$, the scaling center is situated at the origin. With $d\neq0$
the scaling center appears displaced concerning an origin of
coordinates and can be found from the following conditions
\begin{equation}\label{crcyc}
z_c=(\alpha z_{N+1}-z_N)/(\alpha-1),\qquad w_c=(\alpha
w_{N+1}-w_N)/(\alpha-1).
\end{equation}

Let us summarize the algorithm of numerical calculations. At the
first stage, one must define the critical GSK point for the cycles
with not very large periods and dragging the parameter $d$ from
$0$ (critical point of the 1D map) to the necessary value. This
allow to use the known critical GSK point~(\ref{lam}) as a
starting condition at the beginning of the calculation procedure
at $d=0$. Then, with fixed parameter $d$ one must consequently
improve the critical point using the more long-periodic cycles.

Numerical calculations have shown that GSK point for the H\'{e}non
map~(\ref{hen}) with traditional, mentioned in the original work
of H\'{e}non~\cite{Henon} value of parameter $d=-0.3$ is situated
at
\begin{equation}\label{lamd}
\lambda_c^{d=-0.3}=-0.24388583757+0.69478896727i.
\end{equation}
The position of the scaling center is defined by the following
values
\begin{equation}\label{zwc}
z_c=-0.05289397+0.15014466i, \qquad w_c=0.56676843+0.72274977i.
\end{equation}

In the neighborhood of the point~(\ref{lamd}) at the parameter
plane $(\mathrm{Re}\lambda,\mathrm{Im}\lambda)$ the scaling,
characterized by constant $\delta$ takes place (see Fig.~1(b)).
The existence of the infinite number of the cycles with tripling
periods immediately at the critical point $\lambda_c^{d=-0.3}$ is
proved to be true by the structure of the Fourier spectrum. At
Figure~2 the dependence of the spectral intensities versus the
logarithm of frequency $f$ is represented, and amplitudes of
harmonics are rated to the $f^\kappa$, where $\kappa=6.15$ -- is
universal constant, characterizing the self-similar
 fractal structure of the spectrum at the GSK point~\cite{Isaeva3}.

\newpage

\begin{figure}
\centerline{\epsfig{file=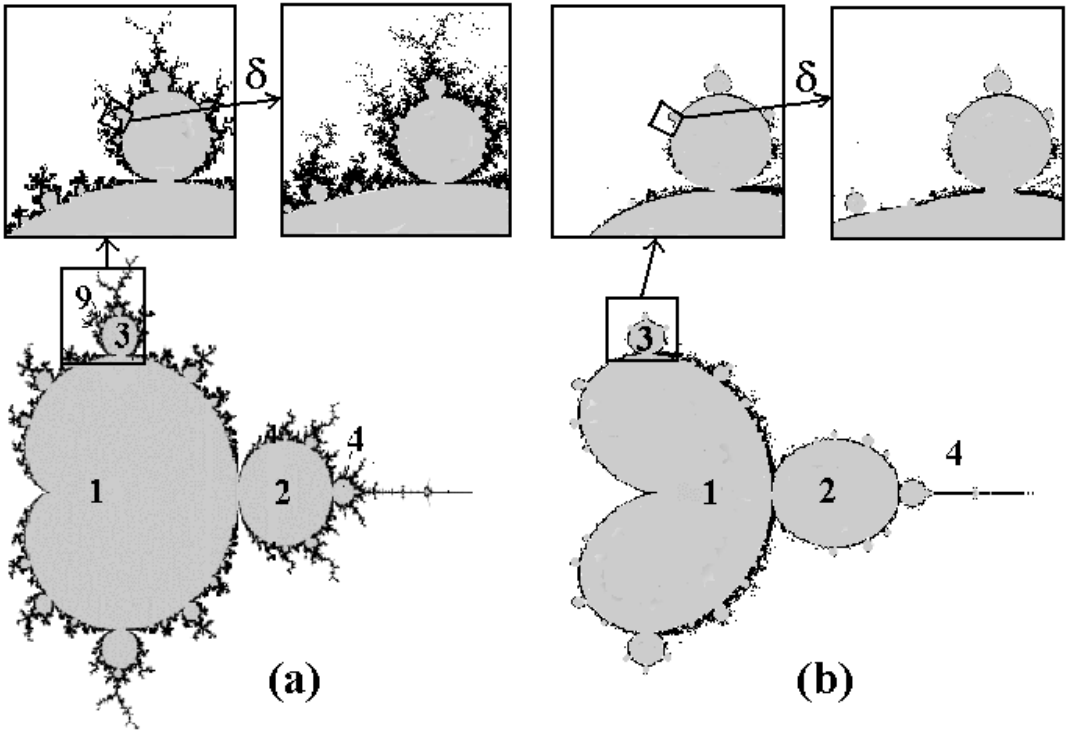,bb = 0 0 420
295,width=0.8\textwidth}}

\caption{Demonstration of the scaling properties of the Mandelbrot
set near the GSK point (critical point is situated at the center
of the small fragments) for the complexified H\'{e}non map~(5) at
the plane of complex parameter $\lambda$ with $d=-0.0$~(a) and
$d=-0.3$~(b). Performed by the multiplication to the complex
scaling constant $\delta$ the fragment of the Mandelbrot set has
the same structure with the previous fragment.}
\end{figure}
\begin{figure}
\centerline{ \epsfig{file=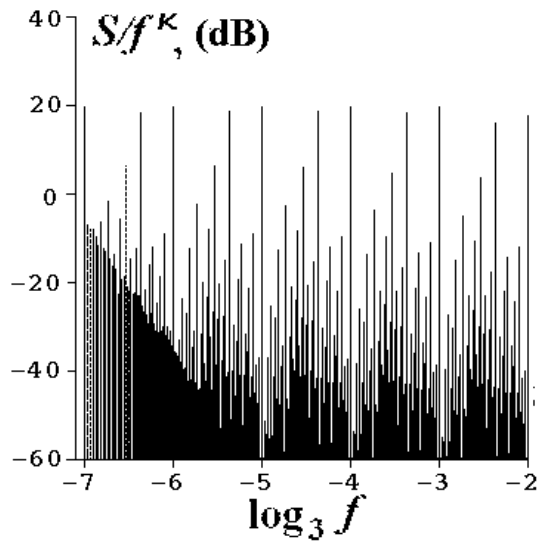,width=0.4\textwidth}}

\caption{Double logarithmic plot of Fourier spectrum of the
signal, originated by the map~(\ref{hen}) at the critical
point~(\ref{lamd}). Harmonics of the tripling frequencies has the
same amplitudes.}
\end{figure}

\end{document}